\documentclass{article}
\usepackage{spconf,amsmath,graphicx,multirow,hyperref,url}
\usepackage{subcaption}
\usepackage{url}

\title{Improving Audio Anomalies Recognition Using Temporal Convolutional Attention Networks}
\name{Qiang Huang ~~ and  ~~Thomas Hain}
\address{ Speech and Hearing Research Group\\
          Department of Computer Science, University of Sheffield\\
\texttt{\{qiang.huang, t.hain\}@sheffield.ac.uk}}
\begin{document}
%
\maketitle
\begin{abstract}
Anomalous audio in speech recordings is often caused by
speaker voice distortion, external noise, or even electric interferences. 
These obstacles have become a serious problem
in some fields, such as high-quality dubbing and speech processing.
In this paper, a novel approach using a temporal convolutional attention network (TCAN) is proposed
to tackle this problem.
The use of temporal conventional network (TCN) can capture long range patterns 
using a hierarchy of temporal convolutional filters.  
To enhance the ability to tackle audio anomalies in different acoustic conditions,
an attention mechanism is used in TCN, where
a self-attention block is added after each temporal convolutional layer.
This aims to highlight the target related features
and mitigate the interferences from irrelevant information.   
To evaluate the performance of the proposed model, audio recordings are collected
from the TIMIT dataset, and are then changed by adding five different types of audio distortions:
gaussian noise, magnitude drift, random dropout, reduction of temporal resolution, and
time warping. Distortions are mixed at different signal-to-noise ratios (SNRs)
(5dB, 10dB, 15dB, 20dB, 25dB, 30dB). 
The experimental results show that the use of proposed model can yield
better classification performances than some strong baseline
methods, such as the LSTM and TCN based models, by approximate 3$\sim$ 10\% relative improvements.

\end{abstract}
\begin{keywords}
Audio anomaly classification, temporal convolutional network, self-attention
\end{keywords}
\section{Introduction}\label{introduction}

Assessing the quality of audio signals is an important consideration in many audio and 
multimedia applications, such as speech recognition, high-quality music recording,
and machine fault detection.
By now, there have been some studies in audio quality assessment \cite{spille2017, fu2018, shah2018, soni2016}.
Fu et al.\cite{fu2018} developed a non-intrusive speech quality evaluation model
to predict PESQ scores using a BLSTM model.
Avila et al., \cite{shah2018} investigated the applicability
of three neural network-based approaches for non-intrusive audio quality assessment
based on mean opinion score(MOS) estimation \cite{Streijl2016}.
These previous studies mainly focused on quality estimations of audio recordings
by predicting a quality score. To our knowledge, few investigated how to identify
what type an anomalous distortion is. Moreover, identifying the type
of audio anomalies will be useful to anomaly detection and audio quality enhancement,
which could benefit various fields in industry.
To tackle the audio anomalies classification, it is highly desirable 
not only to find out whether there exist audio anomalies in audio recordings, 
but also to identify which type an audio distortion belongs to.

In this work, a Temporal Convolutional
Attention Network \cite{Bai2018} is proposed and investigated.
This is because TCNs have advantages in two aspects. Firstly,
TCNs use 1D dilated convolutions \cite{Bai2018_2} to flexibly enlarge their receptive
field through increasing their dilation rate.
This enables TCN to process long-term sequences by 
using a wider part of the input data to contribute to
the output \cite{Oord2016}.
Secondly, unlike the RNN-based methods, its computation is performed layer-wise and
its weights at every time-step are updated simultaneously \cite{Bai2018}. 
By now, there have been already some applications of TCN in activity detection \cite{Lea2017,Farha2019}, 
language processing \cite{Alq2019} and event detection \cite{He2019}.
However, it was found in our experiments that the use of TCN 
did not show satisfying performances on
audio anomalies classification in some conditions, e.g. the SNR
of distortion corrupted signals is relatively low or high. 
For this reason, an attention mechanism is integrated into TCN,
as attention \cite{Vaswani2017, Cheng2016, Parikh2016}
exhibits a better balance between the ability to model long-range dependencies 
and the computational and statistical efficiency. 
Unlike a previous study \cite{Huang2020TCNATTAN} using an attention
block after a TCN block, our work goes deeper into the TCN block
by inserting a self attention layer after each 1D conventional layer. 
The related details of our model will be introduced in the following
sections.

The rest of paper is organised as follows: Section 2 introduces the proposed model architecture in detail.
Section 3 depicts the used dataset and experimental set up. Section 4 presents and analyses
the obtained results, and finally the conclusion and future work are given in Section 5.

\begin{figure*}[htb]
\centering
\begin{minipage}[b]{.28\linewidth}
  \centering
  \centerline{\includegraphics[width=4.6cm, height=3.2cm]{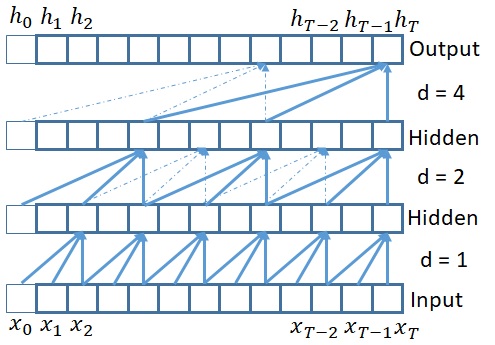}}
  \centerline{(a)}\medskip
\end{minipage}
\begin{minipage}[b]{.28\linewidth}
  \centering
  \centerline{\includegraphics[width=4.0cm, height=3.2cm]{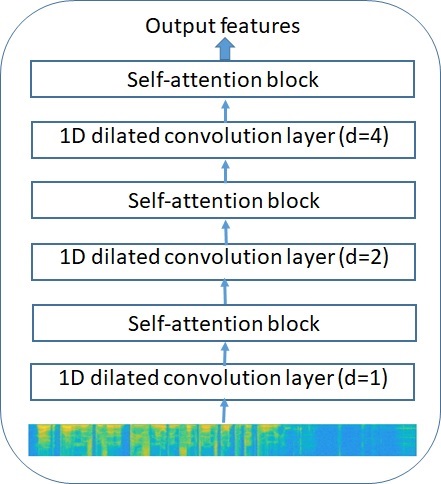}}
  \centerline{(b)}\medskip
\end{minipage}
\begin{minipage}[b]{0.28\linewidth}
  \centering
  \centerline{\includegraphics[width=4.0cm, height=2.8cm]{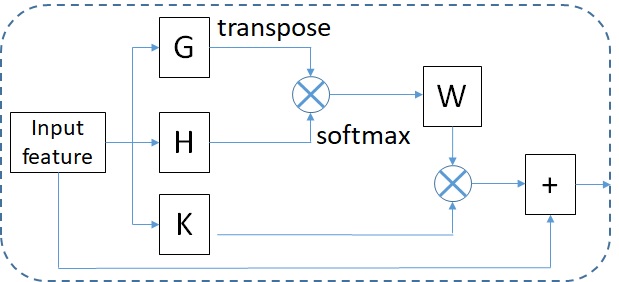}}
  \centerline{(c)}\medskip
\end{minipage}
\hfill
\caption{Architecture of the Temporal Convolutional Attention Network (TCAN): (a) A 3-layer dilated causal convolution; (b) The basic structure of TCAN;
(c) Self-attention block.}
\label{fig:res}
\end{figure*}

\section{Model Architecture}\label{Theoretical Framework}

Given an input audio feature sequence $\textbf{x} = \{x_0, \cdots, x_T\}$,
our aim is to identify the type of audio anomalous distortions $\textbf{A} = \{a_i,\cdots, a_M\}$
in recordings, and $\textbf{M}$ is the number of distortion classes.
As aforementioned in Section 1, the use of TCN is to collect dependencies
over long spans using dilated convolutional layers. 
Figure 1 shows the architecture of proposed approach using the temporal convolutional
attention network. Figure 1(a) illustrates the structure of 
a dilated causal convolution with dilation factors $d = \{1, 2, 4\}$ and filter size $k = 3$.
Figure 1(b) shows the temporal convolutional attention networks
expanded by inserting an attention block between two convolutional layers
of a TCN, and
Figure 1(c) shows the structure a self-attention block.

\subsection{Temporal Convolutional Network}\label{sec:tcn}

As shown in Figure 1(a), the TCN relies on 1D dilated convolutional layers
stacked hierarchically. For a 1D sequence input $\textbf{x} \in \Re^n$ and a filter $f:\{0,\cdots,k-1\} \rightarrow \Re$,
the dilated convolution operation $F$ on element $s$ of the sequence is defined as:
\begin{equation}
  F(s) = \Sigma_{i=0}^{k-1}f(i)\cdot x_{s-d\cdot i}
\end{equation}
where $d$ is the dilation factor, $k$ is the filter size, and $s-d\cdot i$ accounts for the direction of the past\cite{Bai2018}.
It is clear that the TCN is controlled by two parameters, 
dilation factor ($d$) and filter size ($k$).
Dilation is equivalent to introducing a fixed step between every $d$ adjacent
filter taps. Increasing its value can lead to the increment of the depth of the network
(i.e., $d = O(2^i)$ at level $i$ of the network).
When $d = 1$, a dilated convolution reduces to a
regular convolution. Using larger dilation enables an output
at the top level to represent a wider range of inputs, thus
effectively expanding the receptive field of convolutional layers\cite{Bai2018}.
For choosing larger filter sizes $k$, the effective history of one such layer is
$(k-1)d$. 
With above designs, the TCN model is thus able to take similar inputs and produce
similar outputs as RNNs while it is efficient taking advantage of convolution
architectures.

\subsection{Temporal Convolutional Attention Network}\label{sec:tcan}

In comparison with the basic structure of TCN, the TCAN aims to
find which features are more relevant to the recognition target and
which are less or not when search through observed long data streams. 
As shown in figure 1(b), the basic TCN structure is expanded
by adding a self-attention block after a 1D dilated convolutional layer.

The architecture of the used self-attention unit \cite{Purohit2019}
is shown in Figure 1(c).
The input features
are transformed into $G$ and $H$ via 1D convolution, and then generate the attention weights $W$
from $G$ and $H$ by
\begin{equation}
   W = f_{softmax}(G^{T}H)
\end{equation}
where $f_{softmax}$ denotes the softmax function. 
After that, the weighted features $W^{T}K$ are obtained, where
$K$ is another set of features transformed from 1D convolution.
The output features is the sum of the weighted features and
original inputs.

\begin{figure*}[htb]\label{fig:results_accuray}
\centering     
\begin{minipage}[b]{0.31\linewidth}
\centering
\includegraphics[width=56mm, height=36mm]{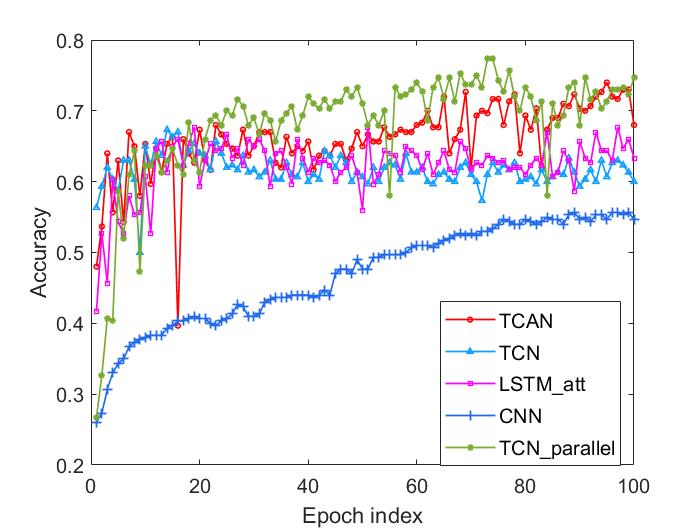}
\centerline{(a) SNR=5dB}
\end{minipage}
\begin{minipage}[b]{0.30\linewidth}
\centering
\includegraphics[width=56mm, height=36mm]{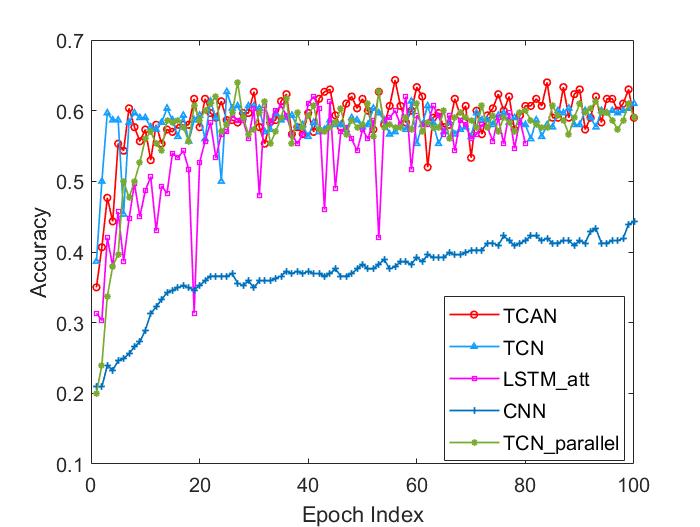}
\centerline{(b) SNR=10dB}
\end{minipage}
\begin{minipage}[b]{0.30\linewidth}
\centering
\includegraphics[width=56mm, height=36mm]{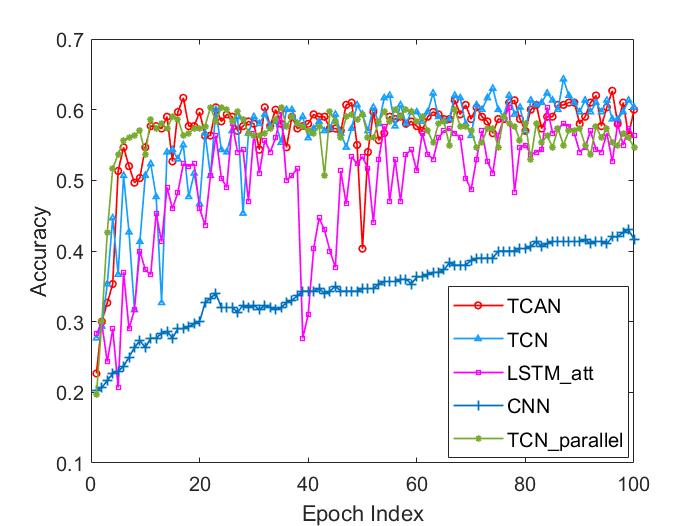}
\centerline{(c) SNR=15dB}
\end{minipage}
\begin{minipage}[b]{0.31\linewidth}
\centering
\includegraphics[width=56mm, height=36mm]{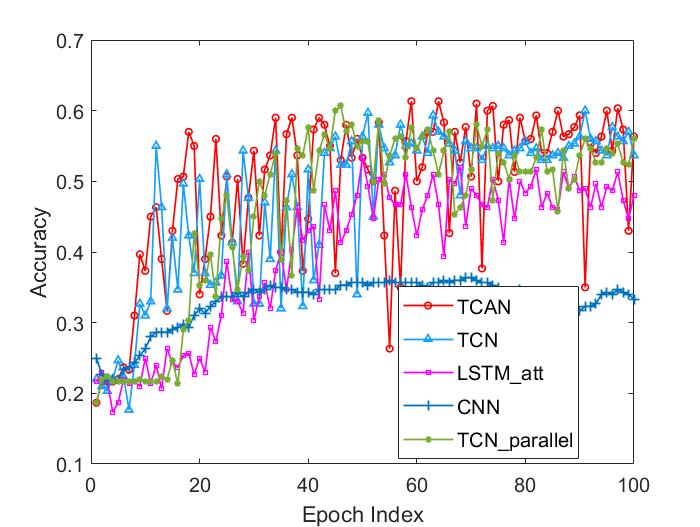}
\centerline{(d) SNR=20dB}
\end{minipage}
\begin{minipage}[b]{0.30\linewidth}
\centering
\includegraphics[width=56mm, height=36mm]{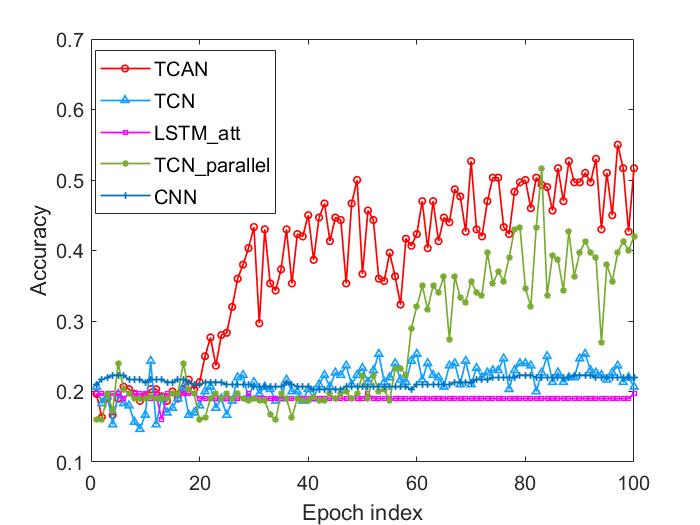}
\centerline{(e) SNR=25dB}
\end{minipage}
\begin{minipage}[b]{0.30\linewidth}
\centering
\includegraphics[width=56mm, height=36mm]{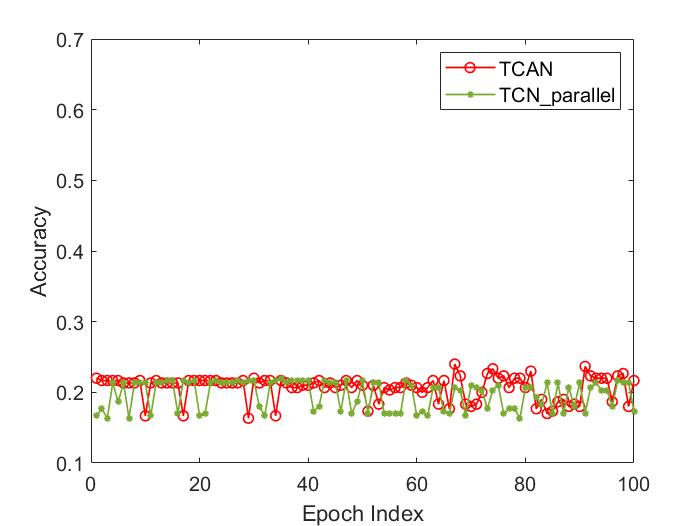}
\centerline{(f) SNR=30dB}
\end{minipage}
\caption{Accuracy comparisons of audio anomalous distortion classification when implementing TCAN and four baseline methods in the condition of different SNRs.}
\vspace{-3mm}
\end{figure*}

\section{Data and Experimental Set Up}\label{sec:data_exp}

\subsection{Data}\label{subsec:data}
In our experiments, the TIMIT dataset \cite{timit92} was used.
3436 recordings, longer than 2.5 seconds, were selected from the original TIMIT training
set and utilized as the training data in this paper.
Meanwhile, 600 recordings selected from the original TIMIT test set in the same
condition were used for evaluation.
The audio anomalous distortion classes were generated 
by changing the original signals in five ways \cite{tsaug},

\begin{description}
\item[Random time warping (Class1):] The time warping is controlled by the number of speed changes and the maximal ratio of max/min speed.
\item[Pooling time series (Class2):] Reduce the temporal resolution without changing the length.
\item[Dropouting values of time series (Class3):] Some random time points in time series are dropped out.
\item[Drifting the value of time series (Class4):] The values of time series are drifted from its original values randomly and smoothly. 
\item[Adding random noise to time series (Class5):] The noise added to every time point of a time series is independent and identically distributed.
\end{description}

As the work in this paper
focuses on the distortions assumed to be caused by speakers,
external natural sounds were not used as noise signals. The four distortion classes
(Class 1 $\sim$ 4) except Class5 are made by only changing the characters of audio signals,
such as temporal resolution and speaking speed. 
In order to evaluate the robustness of the proposed approach,
the recordings corrupted by the five types of distortions are generated at six signal-to-noise ratio (SNR)
levels (5dB, 10dB, 15dB, 20dB, 25dB, 30dB). The larger SNR is, the more difficult it is to identify audio anomalies. 

In all experiments, filter-bank vectors are used to represented input audio features.
2-second audio data is selected from each audio recording and segmented using a 32-ms sliding window with a 16-ms shift.
After conducting a 512-point FFT, each segment is converted into a 40D vector with filter banks.

\subsection{Structure Configuration and Implementation}\label{subsec:config}
In the experiments, the proposed model architecture contains two parts.
The first nine layers consisting of 1D dilated CNN
and attention layers are used to build TCAN, and the last two layers are fully connected
layers used as a classifier.
The dimension of input frame vector
is 40, and 64 kernels ($k=$6) are used in four 1D dilated CNN layers with
$d=\{1,2,4,8\}$.

As a comparison, besides the proposed approach (TCAN),
the related experiments were also conducted 
using another four baseline methods:
TCN$\_$parallel\cite{Huang2020TCNATTAN}, TCN\cite{Bai2018}, CNN\cite{Zhang2017},
and, LSTM$\_$att\cite{Jang2020}.
The method of TCN$\_$parallel connects three temporal convolutional networks (TCNs)
in parallel. Each of TCN is followed by an attention layer and their 
output are then concatenated before sending to a MLP based classifier.
The dilation value (d) and kernel size (k) in TCN$\_$parallel are set to 8 and 6, respectively. 
The TCN baseline is implemented relying on the model structure given in \cite{Bai2018},
where its $d$ and $k$ are set to be 32 and 6, respectively.
The baseline method of LSTM$\_$att makes use of the Bi-directional
LSTM \cite{Gers2002} and an attention mechanisms to capture the long-term dependencies.
The dimensionality of the output space of the LSTM$\_$att is 200.
The baseline method of CNN, originally used for sentence classification, is used 
for audio recording distortion classification, and 
its filter number is set to 64.
 
In all experiments, Adam \cite{Kingma2014} was used as an optimiser
and the initial learning rate was set to 0.001 with 0.95 decay every epoch. 
Classification accuracy is used as a metric in all experiments to evaluate 
performances of the proposed approach and
four baseline methods.

\section{Results Analysis}\label{sec:results}

Figure 2(a)$\sim$2(f) show
audio distortion recognition using the proposed
approach (TCAN) and four baseline methods in the condition
of six SNR. Figure 2(f) only shows the accuracy
curves obtained using TCAN and TCN$\_$parallel when SNR is 30dB. 
This is because
other methods have failed to recognise the type of audio distortions when
SNR is 25dB, and it is thus unnecessary to run them again when SNR is 30dB.
By the first five figures, it can be found that the use of TCAN
can clearly outperform LSTM$\_$att and CNN
in all SNR conditions. This might be related to the following factors.
The CNN relies on kernel size and
focuses more on local dependencies in comparison with the use of TCN.
Although the LSTM takes into account long spans,
the currently observed signals still play relatively important roles
than historical observations.
The structure of TCN tries to mitigate this impact by viewing
all features in a data stream more equally. Moreover,
the use of dilation can further expand the search for longer dependencies,
and the learned information can be passed to a classifier from
bottom to top via a hierarchical structure. 
This might be effective to enable the TCN to capture both long and short term
dependencies of data streams. 

In addition, compared to TCN, the proposed approach can
yield better performances for most SNR cases. The possible reason
is the use of an attention mechanism.
The structure of TCN aims to collect long range features, but
not all the collected features are useful to the task. Some of
them might be irrelevant features and even interferences.
The use of attention mechanism can mitigate this impact by allocating different weights
to target relevant and irrelevant features.

In comparison with TCN$\_$parallel, it seems that TCAN 
works better when SNR is increased. The reason might be
related to where to use attention mechanism. 
As audio distortions become quite weak in the condition of high SNR (e.g. SNR=25dB),
highlighting target relevant information as early as possible might be useful
to mitigate the interferences from strong speech signals.
Following the assumption, the use of TCAN can go deeper than TCN$\_$parallel
to search for target relevant features to reduce possible information loss.

\begin{figure}[t]\label{fig:CM}
\setlength{\belowcaptionskip}{-10pt}
\centering
\includegraphics[width=83mm, height=56mm]{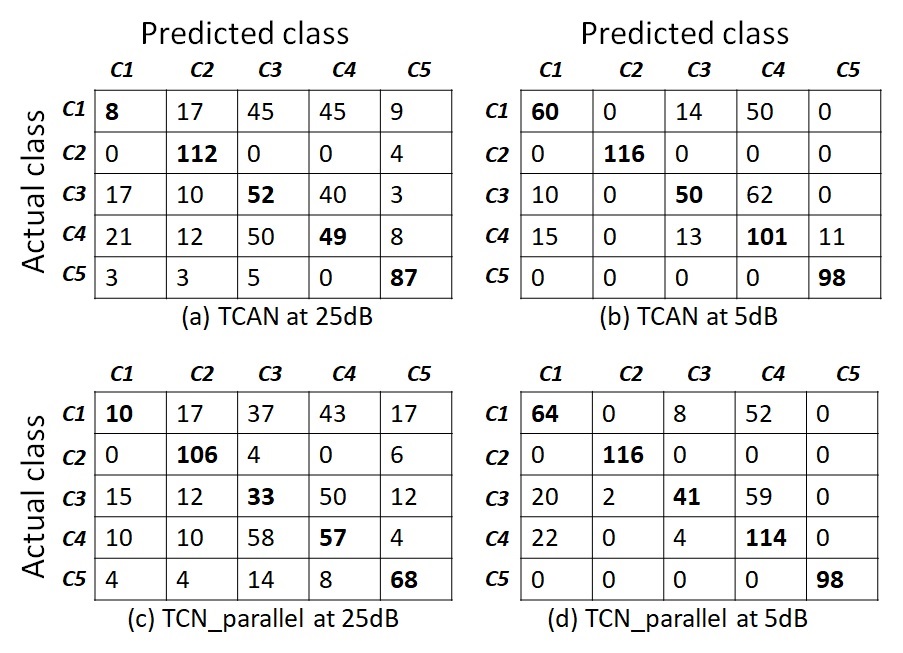}
\caption{Confusion matrix for audio anomalies classification using TCAN and TCN$\_$parallel when SNR is 5dB and 25dB.}
\vspace{-3mm}
\end{figure}

Figure 3 illustrates four confusion matrix tables obtained by using
TCAN and TCN$\_$parallel in the conditions of SNR$=$5dB and SNR$=$25dB, respectively.
In the condition of 5dB,
both methods can yield good and similar recognition performances.
In the condition 25dB, TCAN can better distinguish the five
types of audio distortions than TCN$\_$parallel.  
When SNR is 25dB, the anomalous distortions in recordings are much smaller than those at 5dB. 
This makes class1(C1) be more easily classified as incorrect classes, 
such as class3(C3) and class4(C4), than that at 5dB.
In the same condition, both methods can well recognise two types of audio distortions,
class2(C2) and class4(C4), but
failed to recognise class1(C1).
For the two cases, the possible reason might be related to what parts
of input audio signals are changed. For the audio distortion caused by "random time warping(C1)", 
it has mainly effect on some specific frequencies. This case might also occur within an utterance
if the tune of words or phrases is normally changed by a speaker. This might bring some
extra interferences and make it difficult to learn distinct features
when speech signals are dominant.
For ``pooling time series(C2)'' and ``adding random noise(C5)'', both of them can add distortions on
the whole data sequence, and thus change the feature values in both time and frequency domains simultaneously.
The changes caused by the two types of distortion might be able to bring some salient differences, which
enable them to be found more easily than other distortion classes. 
\begin{table}[h]
  \setlength{\belowcaptionskip}{-10pt}
  \small
  \begin{center}
    \label{tab:dilation}
    \begin{tabular}{|c|c|c|c|c|c|c|}
      \hline
      \textbf{d=1} & \textbf{d=2} & \textbf{d=4} & \textbf{d=8} & \textbf{d=16} & \textbf{d=32} & \textbf{d=64} \\\hline
      48.6 & 54.3& 52.3& 53.6& 55.3& 56.3 & 52.3 \\
      \hline
    \end{tabular}
    \caption{Classification accuracy (\%) with different dilation values (SNR=25dB, k=6).}
  \end{center}
  \vspace{-6mm}
\end{table}

Table 1 shows the change of classification accuracy when different dilation (d) is set.
It seems that there is a weak tendency that accuracy is slightly better when increasing $d$.
The possible reason might be larger $d$ is able make the model learn information from a longer range.

\vspace{-6mm}
\section{Conclusion and Future Work}\label{sec:conclusion}
A novel structure for audio distortion classification was designed by
using the temporal convolutional attention network (TCAN).
It can well distinguish five different
types of audio distortions in condition of different SNRs. 
Moreover, the obtained results have
shown its robustness to in comparison with several strong
baseline methods, especially 
when SNR is relatively low (5dB) or high (25dB).

In future, work in three aspects will be taken into account.
Firstly, some advanced neural network technologies
will be used to assess audio quality.
Secondly, the classification technologies will be evaluated on
large-sized speech datasets and in various acoustic conditions.
Thirdly, more audio distortion types and model configurations will be also
evaluated to make the system work in some practical applications.

~~~~~~~~~~~~~~~~~~~~~~~~~~\textbf{Acknowledgements}\\
This work was supported by Innovate UK Grant number 104264 MAUDIE.

\bibliographystyle{IEEEbib}
\bibliography{strings,refs}

\end{document}